\documentclass[DIV=calc,%
							paper=a4,%
							fontsize=11pt,%
							twocolumn]{scrartcl}	 					

\pdfoutput=1
\pdfmapfile{+txfonts.map}

\usepackage[english]{babel}										
\usepackage[protrusion=true,expansion=true]{microtype}				
\usepackage{amsmath,amsfonts,amsthm}					
\usepackage[final]{graphicx}									
\usepackage{xcolor}									
\usepackage[normal, small, hypcap, up, labelfont=bf, textfont=it]{caption}	
\usepackage{epstopdf}												
\usepackage{subfig}													
\usepackage{booktabs}												
\usepackage{fix-cm}													
\usepackage{amssymb,amsfonts}       
\usepackage{dsfont}
\usepackage{bbm}
\usepackage{pstricks}
\usepackage{cite}
\usepackage[utf8]{inputenc}
\usepackage[perpage,symbol*]{footmisc}
\usepackage{pstricks}
\usepackage[varg]{txfonts}
\usepackage{balance}
\usepackage{fancyhdr}	
\usepackage{hyperref}
\usepackage{verbatim}
\usepackage{fontenc}

\DeclareCaptionFont{mycolor}{\color[HTML]{0000FF}}
\usepackage{sectsty}													
\allsectionsfont{
	\color[HTML]{31ADF3}\usefont{OT1}{phv}{b}{n}
	}

\sectionfont{
	\color[HTML]{31ADF3}\usefont{OT1}{phv}{b}{n}
	}

\usepackage{fancyhdr}												
\usepackage{lettrine}
\newcommand{\initial}[1]{%
     \lettrine[lines=3,lhang=0.3,nindent=0em]{
     				\color[HTML]{31ADF3}
     				{\textsf{#1}}}{}}

\usepackage{titling}															

\newcommand{\HorRule}{\color[HTML]{31ADF3}
									  	\rule{\linewidth}{1pt}%
										}

\pretitle{\vspace{-30pt} \begin{flushleft} \HorRule 
				\fontsize{34}{34} \usefont{OT1}{phv}{b}{n} \color[HTML]{31ADF3} \selectfont 
				}
\title{Is Schr\"{o}dinger's Cat Alive?}					
\posttitle{\par\end{flushleft}\vskip 0.5em}

\preauthor{\begin{flushleft}\large \lineskip 0.5em \usefont{OT1}{phv}{b}{sl} \color[HTML]{31ADF3}}
\author{Mani L. Bhaumik\\[8pt]}											
\postauthor{\footnotesize \usefont{OT1}{phv}{m}{sl} \color[HTML]{000000} 
Department of Physics and Astronomy, University of California, Los
Angeles, USA. E-mail: \href{mailto:bhaumik@physics.ucla.edu}{bhaumik@physics.ucla.edu}\\[10pt]		
\scriptsize\usefont{OT1}{phv}{m}{n} \color[HTML]{31ADF3}{\textbf{Editors: \emph{Zvi Bern} \& \emph{Danko Georgiev}} }\\[5pt]
\color[HTML]{000000}{Article history: Submitted on December 2, 2017;  Accepted on December 9, 2017; Published on December 11, 2017.}
					\par\end{flushleft}\HorRule}

\date{}																				

\begin{document}
\maketitle
\thispagestyle{fancy} 			
\initial{E}\textbf{rwin Schr\"{o}dinger is famous for presenting his wave equation of motion
that jump-started quantum mechanics. His disenchantment with the Copenhagen
interpretation of quantum mechanics led him to unveil the Schr\"{o}dinger's
cat paradox, which did not get much attention for nearly half a century.
In the meantime, disappointment with quantum mechanics turned his
interest to biology facilitating, albeit in a peripheral way, the
revelation of the structure of DNA. Interest in Schr\"{o}dinger's cat
has recently come roaring back making its appearance conspicuously
in numerous scientific articles. From the arguments presented here,
it would appear that the legendary Schr\"{o}dinger's cat is here to stay,
symbolizing a profound truth that quantum reality exists at all scales;
but we do not observe it in our daily macroscopic world as it is masked
for all practical purposes, most likely by environmental decoherence
with irreversible thermal effects.\\ Quanta 2017; 6: 70--80.}
\begin{figure}[b!]
\rule{245 pt}{0.5 pt}\\[3pt]
\raisebox{-0.2\height}{\includegraphics[width=5mm]{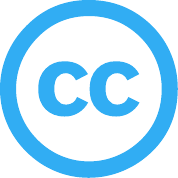}}\raisebox{-0.2\height}{\includegraphics[width=5mm]{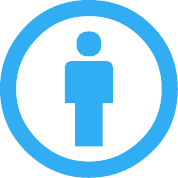}}
\footnotesize{This is an open access article distributed under the terms of the Creative Commons Attribution License \href{http://creativecommons.org/licenses/by/3.0/}{CC-BY-3.0}, which permits unrestricted use, distribution, and reproduction in any medium, provided the original author and source are credited.}
\end{figure}
\section{Introduction}

Like most cats, this one slipped into the room practically unnoticed.
Schr\"{o}dinger's cat \cite{Schrodinger1935a,Schrodinger1935b,Schrodinger1935c,Schrodinger1935d,Schrodinger1935e}, which is supposedly both alive and dead
at the same time, has by now attained a celebrated status in both
scientific and public spheres. In addition to its surprisingly frequent
mention in quantum physics articles, the quantum computation community
has bestowed it a permanent berth by coining a cat state in its repute.
The appearance of a cat's image adorning the jacket of a popular undergraduate
textbook \cite{Griffiths2004} as well as two popular science books on the subject \cite{Gribbin1984,Rowlands2015}
speaks volumes about its privileged place in scientists' fancy. It
has come to be the common metaphor for superposition so ubiquitous
in the quantum domain. Perhaps because of its somewhat melodramatic
character, it has also secured a place in popular culture
with appearances in literature, television, film, music, cartoons,
jokes, and video games. It would be rather fascinating to follow the
trail of this captivating story since in a real sense Schr\"{o}dinger's
cat is a conspicuous symbolic entity representing a profound reality,
namely: the manifest coexistence of the quantum and classical reality,
which appear so particularly different in their attributes.

At the very dawn of the twentieth century, Max Planck initiated the
quantum revolution by introducing the idea of a quantum \cite{Planck1901}. After years
of frustrating failure to work out an accurate formula for thermal
radiation of a black body, Planck postulated that radiation energy
is emitted and absorbed only in discrete packages he called quanta.
But he was, however, a reluctant revolutionary and did not believe
in the physical existence of the quantum, which so uniquely contributed
in formulating the correct equation for the black body radiation.

It was Albert Einstein who recognized the reality of a quantum and
suggested that radiation in space also consists of discrete quanta \cite{Einstein1905}.
With near unanimity, the entire physics community---including Planck
himself---remained very skeptical about the postulate since by then
Maxwell's theory of electromagnetic radiation in terms of continuous
wave motion had, through its numerous confirmations, become ingrained
in people's minds. It also evoked the counter intuitive idea of a
wave-particle duality. Nevertheless, Einstein unwaveringly persisted
almost for two decades as the sole champion keeping the nascent quantum
revolution alive. His tenacity was finally vindicated spectacularly
by the discovery of the Compton effect in 1923, which demonstrated
that X-rays could be deflected like billiard balls by an electron \cite{Compton1923}.

With an overwhelming support from Einstein, the young French graduate
student, Louis de Broglie galvanized the quantum revolution in 1924
by boldly postulating in his doctoral thesis that even matter particles
could have wave like properties. His hypothesis gained support from
the fact that it could explain the radius of the stationary orbits
of the electron in Niels Bohr's audacious model of the hydrogen atom,
offered in 1913 \cite{Bohr1,Bohr2,Bohr3}. The radius of a Bohr's conjectured orbit was an integral
multiple of de Broglie's proposed wavelength of the electron. Years
later, Einstein confided that he also came up with the idea of matter
waves but did not publish it due to a lack of any evidence. After
de Broglie submitted his thesis, Einstein ardently appealed to the
physicists to look for an experimental proof of the postulated waves.
Soon Clinton Davisson and Lester Germer provided the evidence in 1927
by observing a diffraction pattern in the beam of electrons scattered
by nickel crystals \cite{Davisson1927}.

In the meantime, through his correspondence with Einstein \cite[p.~412]{Mehra1987},
Erwin Schr\"{o}dinger became fascinated by the radical concept of de Broglie's
matter wave. At the request of Peter Debye, Schr\"{o}dinger gave a seminar
at ETH in Zurich toward the end of November 1925, enthusiastically
explaining how the matter waves gave support to Niels Bohr's atomic
model with stable electronic orbits. In the seminar, Debye expressed
some skepticism about how waves, which normally spreads out, could
confine itself in stable atomic orbits. He suggested that a relevant
wave equation of motion should be formulated to deal with a wave in
a proper way \cite[pp.~419--421]{Mehra1987}.

Schr\"{o}dinger took up the challenge, since apparently he was already
thinking about the subject himself, and promptly formulated his groundbreaking
equation of motion for matter waves in early 1926 \cite{Schrodinger1926a,Schrodinger1928}. With some
assistance from
his associate, mathematical physicist
Hermann Weyl, Schr\"{o}dinger found the solutions for the standing waves
of the discrete orbits of the hydrogen atom \cite[p.~1165]{Mehra1987b}. Schr\"{o}dinger stated in
his first paper
\begin{quote}
The essential thing seems to me
to be, that the postulation of ``whole numbers''
no longer enters into the quantum rules mysteriously, but that we
have traced the matter a step further back, and found the ``integralness''
to have its origin in the finiteness and single-valuedness of a certain
space function. \cite[p.~9]{Schrodinger1928}
\end{quote}

The astounding success of his
equation in giving precise explanation of the quantized orbital motion
of an electron in a hydrogen atom initiated a storm of activities
and the quantum revolution went into full swing.

\section{Trail to the Schr\"{o}dinger Equation}

As we have just described, after the demonstration of the Compton
effect, wave-particle duality of massless particles like photons was
established beyond a reasonable doubt. Louis de Broglie's profound
insight to extend the idea to massive particles was a feat of genius.
He correctly surmised that a massive particle like an electron would
have a characteristic internal frequency $\nu_{0}$ in compliance
with the Planck--Einstein relation
\begin{equation}
E=m_{0}c^{2}=h\nu_{0}
\end{equation}
that would obey the rules of special relativity to provide the length
of the matter wave $\lambda=h/p$ or equivalently $p=\hbar k$ analogous
to that of the photon. However, he struggled with a derivation of
the matter wave using the special theory of relativity to such an
extent that he portrayed the matter wave to be merely a fictitious
guiding wave to be used for the kinematics of the matter particle.
Consequently, the matter wave proposition received the moniker de
Broglie's hypothesis instead of a theory. We now know that the relationship
$p=\hbar k$ can be elegantly comprehended using the four-vector procedure
of the special theory of relativity.

In an earlier presentation \cite{Bhaumik2015}, it was shown that the wave packet
of a quantum particle could be deduced from an objective reality to
be given by
\begin{equation}
\psi(x,t)=\frac{1}{\sqrt{2\pi}}\int_{-\infty}^{\infty}\tilde{\psi}(k)e^{\imath(kx-\omega_{k}t)}dk\label{eq:2}
\end{equation}
The group velocity of the wave-packet for a massless particle like
a photon is
\begin{equation}
v_{g}=\frac{d\omega}{dk}=\frac{d(kc)}{dk}=c
\end{equation}
For a massive particle like an electron, the group velocity of the
wave packet in Eq.~(\ref{eq:2}) is no longer equal to the constant
$c$, but a variable velocity $v$ depending upon its energy-momentum.
Accordingly, the spacetime dependence $e^{\imath(kx-\omega_{k}t)}$
must change frequently, which can be deduced from the special theory
of relativity.

The transformation of the momentum four-vector $(\frac{E}{c},\vec{p})$
and the wave four-vector $(\frac{\omega}{c},\vec{k})$ that keeps
their magnitude invariant is the Lorentz transformation. Considering
a laboratory frame $S$ and a rest frame $S^{\prime}$ with a boost
velocity $v$ in the $x$ direction, the Lorentz transformation relations
for the momentum four-vector and the wave four-vector are
\begin{eqnarray}
\frac{E^{\prime}}{c} & = & \gamma(\frac{E}{c}-\beta p_{x})\\
p_{x}^{\prime} & = & \gamma(p_{x}-\beta\frac{E}{c})\\
p_{y}^{\prime} & = & p_{y}\\
p_{z}^{\prime} & = & p_{z}
\end{eqnarray}
and
\begin{eqnarray}
\frac{\omega^{\prime}}{c} & = & \gamma(\frac{\omega}{c}-\beta k_{x}) \\
k_{x}^{\prime} & = & \gamma(k_{x}-\beta\frac{\omega}{c}) \\
k_{y}^{\prime} & = & k_{y} \\
k_{z}^{\prime} & = & k_{z}
\end{eqnarray}
where $\beta=\frac{v}{c}$ and $\gamma=1/\sqrt{1-\beta^{2}}$.

To determine the proportionality constant between the two four-vectors,
we correlate their timelike components after multiplying the component
of the wave four-vector by $\hbar$:
\begin{eqnarray}
\frac{E^{\prime}}{c} & = & \gamma(\frac{E}{c}-\beta p_{x})\label{eq:5}\\
\frac{\hbar\omega^{\prime}}{c} & = & \gamma(\frac{\hbar\omega}{c}-\beta\hbar k_{x})\label{eq:6}
\end{eqnarray}
Subtracting Eq. (\ref{eq:6}) from Eq. (\ref{eq:5}), we find
\begin{equation}
\frac{E^{\prime}}{c}-\frac{\hbar\omega^{\prime}}{c}=\gamma(\frac{E}{c}-\frac{\hbar\omega}{c}-\beta p_{x}+\beta\hbar k_{x})\label{eq:7}
\end{equation}
Since the laws of physics are the same in all inertial frames of reference,
the Planck's law $E=\hbar\omega$ in frame $S$ should hold true in
frame $S^{\prime}$, giving us $E^{\prime}=\hbar\omega^{\prime}$.
Thus Eq. (\ref{eq:7}) reduces to
\begin{equation}
\gamma\beta(p_{x}-\hbar k_{x})=0
\end{equation}
According to the zero product property of algebra, either $\gamma\beta=0$
or $p_{x}-\hbar k_{x}=0$. Because $\gamma\beta$ is non-zero, we
obtain 
\begin{equation}
p_{x}=\hbar k_{x}\label{eq:9}
\end{equation}
Or more generally, $p=\hbar k$ irrespective of the mass of the particle,
zero or otherwise. This relationship for a massive particle is the
celebrated de Broglie hypothesis. However, we now realize that the
relationship can indeed be derived from fundamental considerations
and does not need to be a mere hypothesis.

As presented earlier \cite{Bhaumik2016}, using the relationship in Eq.~(9) the
group velocity of the wave packet is $v_{g}=v$, which is the velocity
of the ``particle'' represented
by the wave packet. Therefore, the wave packet moves with the velocity
of either a massive or a massless particle. Steven Weinberg also confirms
this result using a slightly different consideration \cite[pp.~11--12]{Weinberg2013}.

\section{Kinematics of the wave packet}

Following Schr\"{o}dinger, the equation of motion for the wave packet
given in Eq. (\ref{eq:2}) can now be formulated using the relationship
in Eq. (\ref{eq:9}).

Schr\"{o}dinger's astute intuition was to express the phase of a plane
wave utilizing Eq. (\ref{eq:9}) to achieve
\begin{equation}
\Psi(x,t)=Ae^{\imath(kx-\omega t)}=Ae^{\frac{\imath}{\hbar}(px-Et)}\label{eq:10}
\end{equation}
and to realize that the first order partial derivatives were:

\noindent with respect to space
\begin{equation}
\frac{\partial\Psi}{\partial x}=\frac{\imath}{\hbar}pAe^{\frac{\imath}{\hbar}(px-Et)}=\frac{\imath}{\hbar}p\Psi
\end{equation}
or
\begin{equation}
p\Psi=-\imath\hbar\frac{\partial\Psi}{\partial x}
\end{equation}
thus the momentum operator is
\begin{equation}
\hat{p}=-\imath\hbar\frac{\partial}{\partial x}
\end{equation}
and with respect to time
\begin{equation}
\frac{\partial\Psi}{\partial t}=-\frac{\imath}{\hbar}EAe^{\frac{\imath}{\hbar}(px-Et)}=-\frac{\imath}{\hbar}E\Psi
\end{equation}
or 
\begin{equation}
E\Psi=\imath\hbar\frac{\partial\Psi}{\partial t}\label{eq:15}
\end{equation}
thus the energy operator is
\begin{equation}
\hat{E}=\imath\hbar\frac{\partial}{\partial t}
\end{equation}
for any state of energy $E$.

Max Born generalized this relationship for any system described by
a Hamiltonian $\hat{H}$. The time dependence of any wave function,
whether or not for a state of definite energy, is then given by
\begin{equation}
\imath\hbar\frac{\partial\Psi}{\partial t}=\hat{H}\Psi\label{eq:16}
\end{equation}

The equation~(\ref{eq:16}) is known as the general Schr\"{o}dinger equation.
It is indeed quite general and used throughout quantum
mechanics, for everything from the Dirac equation to quantum field
theory, by utilizing various complicated expressions for the Hamiltonian $\hat{H}$.

If the Hamiltonian itself is not explicitly dependent on time, Eq.
(\ref{eq:15}) shows that
\begin{equation}
\hat{H}\Psi=E\Psi
\end{equation}
For a single non-relativistic particle of mass $m$, its energy in
a potential $V(x)$ is $\frac{p^{2}}{2m}+V(x)$ giving us
\begin{equation}
-\frac{\hbar^{2}}{2m}\nabla^{2}\Psi(x)+V(x)\Psi(x)=E\Psi(x)\label{eq:18}
\end{equation}
This is the time-independent Schr\"{o}dinger equation for a single non-relativistic
particle of energy $E$. The equation~(\ref{eq:18}) solved for the
hydrogen atom, with the three boundary conditions that: $\Psi(x)$ is single-valued,
$\Psi(x)$ returns to the same value if $x$ goes around a closed
curve, and $\Psi(x)$ vanishes as the magnitude of $x$ goes to infinity,
accurately reproduced Bohr's formula for the discrete energy levels
but without any assumption.

Einstein wholeheartedly endorsed the paper since he considered the
wave nature of particle gave a clearer intuitive picture as compared
to the rather abstract matrix mechanics introduced by Werner Heisenberg,
Max Born and Pasqual Jordan in 1925. In rapid succession within a
few months after his first paper, Schr\"{o}dinger was able to demonstrate
that the analytical solution of his time independent wave equation
indeed predicts the discrete energy levels of several other
non-relativistic quantum systems, among them, the quantum harmonic oscillator,
diatomic molecule, and Stark effect \cite{Schrodinger1928,Schrodinger1926b}. With Einstein's commendation
and the immense success of the wave equation, Schr\"{o}dinger equation
became the benchmark for wave mechanics at the time bestowing its
inventor an instant renown. The very following year, he was honored
with the venerable chair formerly held by Max Planck at the Friedrich
Wilhelm University in Berlin.

After the spectacular success of the time independent wave equation,
Schr\"{o}dinger focused his attention on developing a time dependent equation
for treating problems in which the quantum system changes with time
as in scattering problems. Using the total energy $H=\frac{p^{2}}{2m}+V(x)$
for a single non-relativistic particle in a potential $V(x)$, and
taking the square of the momentum operator, Schr\"{o}dinger soon unveiled
his time dependent equation of motion for a wave packet
\begin{equation}
\imath\hbar\frac{\partial\Psi(x,t)}{\partial t}=\hat{H}\Psi(x,t)=-\frac{\hbar^{2}}{2m}\nabla^{2}\Psi(x,t)+V(x)\Psi(x,t)\label{eq:19}
\end{equation}
Clearly, the plane wave in Eq. (\ref{eq:10}) is a solution of the
Schr\"{o}dinger equation~(\ref{eq:19}). However, the Schr\"{o}dinger equation
is a linear differential equation. So a linear combination of plane
waves is also a valid solution. The wave packet in Eq. (\ref{eq:2})
is just such a linear combination. The momentum wave function $\tilde{\psi}(k)$
appearing in the integrand is an integral of the position wave functions
since the position and momentum space wave functions are Fourier transforms
of each other. Therefore the kinematics of the wave packet described
in Eq.~(\ref{eq:2}) can be processed using the Schr\"{o}dinger equation
(\ref{eq:19}).

This distinctive aspect of superposition of solutions of the linear
Schr\"{o}dinger equation is its unique signature. It should not be surprising
since in our daily lives, we do see water wavelets superpose when
they come in contact. In the quantum domain as well, we see chargeless bosons like photons and gravitons exhibit superposition by producing
macroscopic classical waves. But the fermion wave functions, being
antisymmetric, in addition to some of them having different charges,
are not alike and do not produce classical waves. However, the bizarre
implication of superposition, especially as construed by the Copenhagen
interpretation for measurement, made both Einstein and Schr\"{o}dinger,
the pioneering leaders of the quantum revolution, enormously unhappy.

\section{Probability Amplitude}

In his initial studies involving the time independent wave equation
to predict energies of discrete quantum states, Schr\"{o}dinger did not
have to be too concerned about what exactly the wave function $\Psi(x,t)$
did represent. It would be tempting to think, as in fact Schr\"{o}dinger
originally did, that the wave function represented a smeared out charge
distribution of the electron. In an experiment involving the scattering
of electrons, Max Born realized that Schr\"{o}dinger's contention that
the wave function represented a charge distribution could not be sustained
\cite{Born1926}. Instead he suggested as it appears in his Nobel lecture
\begin{quote}
Again an idea of Einstein's gave me the lead. He had tried to make
the duality of particles---light quanta or photons---and waves comprehensible
by interpreting the square of the optical wave amplitudes as probability
density for the occurrence of photons. This concept could at once
be carried over to the $\psi$-function: $|\psi|^{2}$~ought to represent
the probability density for electrons (or other particles). \cite{Born1954}
\end{quote}
This is known as the Born's rule.

But recalling how the wave packet had come to be \cite{Bhaumik2015}, it would
have been obvious that Born was correct. The wave packet consists
of irregular disturbances, the sum total of which represents the mass,
energy momentum, charge of a particle like electron. Therefore, the
wave function is in fact a function of probability amplitude for finding
the particle.

It should be noted, however, that although the attributes of the various
irregular disturbances are mostly characteristics of their respective
quantum fields with different charge, spin, etc., they have one aspect
in common. The element of disturbance in energy is identical for all
fields. Energy density of a wave is given by the square of its amplitude.
Therefore, to get the probability density, we have to take the square
of the amplitude of the wave function, which usually involves a complex
quantity. Consequently, the square of the amplitude $\Psi^{*}(x,t)\Psi(x,t)$,
which is the probability density function $P(x,t)$, should represent
the probability density for finding a particle in position space at
time $t$. As discussed earlier, this is acknowledged as the renowned
Born's rule, which is a necessary hypothesis of quantum mechanics.
But as we have just presented, it is a natural consequence of the
reality of the wave function revealed in a previous article \cite{Bhaumik2015}
and does not need to be a hypothetical rule.

However, it is of critical importance that the wave function is normalized
\begin{equation}
\int_{-\infty}^{\infty}\Psi^{*}(x,t)\Psi(x,t)dx=1
\end{equation}
since the equation describes one particle with the sum of probability
to be $1$. Born's probabilistic description of the wave function
was promptly taken over by Niels Bohr in Copenhagen who in essence
became the father of the long-reigning Copenhagen interpretation of
quantum mechanics.

According to the Copenhagen interpretation, physical systems in superposed
states do not possess real properties prior to being measured, and
quantum mechanics can only predict the probabilities that measurements
will produce certain results. The act of measurement using a macroscopic
device by an observer collapses the wave function, causing the set
of probabilities to reduce to only one of the possible values.

It is well known that both pioneers of quantum mechanics, Einstein
and Schr\"{o}dinger, were extremely troubled by the Copenhagen interpretation.
Much has been written on the famous Bohr--Einstein debates. The concerns
of Einstein have been discussed in greater detail in an earlier communication
\cite{Bhaumik2015}. His most strident objection was the almost cult-like denial
by the proponents of the Copenhagen interpretation of any reality
in a quantum system prior to measurement. Among others, in one noteworthy
sarcastic comment he states
\begin{quote}
The present quantum theory is unable to provide the description of a real state of physical facts, but only of an (incomplete) knowledge of such. Moreover, the very concept of a real factual state is debarred by
the `orthodox' quantum theoreticians. The situation arrived at corresponds almost
exactly to that of the good old Bishop Berkeley. \cite[pp.~73--74]{Jammer1982}
\end{quote}
He expected the proponents to be at least open, as he himself was,
to the possibility that quantum mechanics as currently formulated
might be incomplete.

Schr\"{o}dinger's main objection appears to be the rather fictional depiction
of the superposed states and their probability interpretation as well
as the collapse of the wave function. He envisioned that after a unitary
evolution of his wave equation, it could be somehow extrapolated up
to the macroscopic scale of the measuring device. Sometime later,
he sarcastically commented that
\begin{quote}
I don't like it, and I'm sorry I ever had anything to do with it.
\cite[p.~v]{Gribbin1984}
\end{quote}

\section{Quantum Entanglement}

Einstein expressed his dismay regarding the implications of quantum
mechanics very publicly through the famous Bohr--Einstein debates
beginning at the Solvay conference in 1927, of which Bohr was the
generally supposed winner. Bohr, meanwhile, was troubled by none of
the elements that appalled Einstein. He made his own peace with the
contradictions by asserting a principle of complementarity that emphasized
the role of the observer over the observed and placed limits on what
the observer could know about a quantum system. Given his disenchantment
with the Copenhagen interpretation, Einstein's attention shifted from
quantum mechanics primarily toward his interest in developing a unified
theory of gravity and electro-magnetism. After the well-known turbulent
times in Germany, Einstein left Europe and in 1933, permanently settled
down at the Institute for Advanced Study in Princeton, where he finally
had the time and serenity to resume his thoughts on quantum mechanics.

In 1935, Albert Einstein together with Boris Podolsky and Nathan Rosen published
the seminal EPR paper in an attempt to show that the description of
reality as given by a wave function was not complete \cite{EPR}. Einstein tried
to illustrate that superposition of the quantum states of two particles
cannot be real. Because when the particles are separated by an arbitrarily
large distance, measuring a particular property of one of the particles
would instantly affect the other.
In a letter to Born, Einstein wrote
\begin{quote}
I admit, of course, that there is a considerable amount of validity
in the statistical approach which you were the first to recognize
clearly as necessary given the framework of the existing formalism.
I cannot seriously believe in it because the theory cannot be reconciled
with the idea that physics should represent a reality in time and
space, free from spooky actions at a distance. \cite[p.~158]{Einstein1971}
\end{quote}
Because of Einstein's repute, the EPR article generated some impressive
newspaper headlines. But the physics community essentially ignored
it for quite a while. Later, in the 1950s, David Bohm tried to show
that Einstein's desired element of reality could be supported by existence
of some local hidden variables. Bohm's ideas met with stiff resistance
from the scientists.

It was not until 1964, when John Bell started his seminal analysis
known as Bell's inequality that helped to discard the existence of
local hidden variables, proposed by Bohm, but indicated the existence
of nonlocality revealed in Bohm's investigation. Bell ingenuously
proposed an approach to experimentally verify nonlocality between
a pair of quantum particles in superposed states, which are now called
entangled states as coined by Schr\"{o}dinger. Numerous experimental verifications
of the existence of quantum entanglement beyond any reasonable doubt
have now been accomplished establishing that quantum superposition
is undoubtedly real. Using a satellite, Chinese scientists
have demonstrated entanglement over a distance as large as 1200 kilometers
patently attesting that Einstein's ``spooky action
at a distance'' is here to stay \cite{Yin2017}. He still might
have derived some satisfaction from knowing that in conformity with
his special theory of relativity, no meaningful signal can be sent
using entanglement and the spookiness of the action is circumscribed.

As explained by Narnhofer and Thirring, in quantum field
theory almost everything is entangled in the quantum and the mesoscopic
domains \cite{Narnhofer2012}. Penrose finds this extremely puzzling, stating
\begin{quote}
Since, according to quantum mechanics, entanglement is such a ubiquitous
phenomenon---and we recall that the stupendous majority of quantum
states are actually entangled ones---why is it something that we barely
notice in our direct experience in the world? \cite[p.~591]{Penrose2004}
\end{quote}
It is rather ironic that the very pivotal EPR thought experiment in
which Einstein attempted in effect to show the impossibility of quantum
entanglement would lead to one of the most astounding discoveries
of the twentieth century. It has opened a floodgate of activities
in quantum physics. The possible uses of quantum entanglement in a
variety of novel applications such as quantum cryptography, quantum
computation, and quantum teleportation have become areas of very active
research.

It has also provided an immense stimulus to basic scientific investigations.
Experts such as Maldacena and Susskind postulate that ER=EPR
implying there is an as yet unknown quantum mechanical version of
a classical wormhole that permits quantum entanglement \cite{Maldacena2013}. There is a
possibility that the quantum fluctuations of the fields are themselves
entangled facilitating a quantum mechanical Einstein--Rosen bridge \cite{Bhaumik2014}. Most
exciting perhaps is the intriguing prospect that spacetime
itself could be stitched together by entanglement \cite{vanRaamsdonk2010,Swingle2012}.

\section{Arrival of the Schr\"{o}dinger's Cat}

Like Einstein, Schr\"{o}dinger fled the rapidly deteriorating political
situation and left Berlin in 1933 for England to join the University
of Oxford. Soon after he arrived, he received the Nobel Prize, which
should have augured a very comfortable existence for him. Instead,
the glare that focused on his personal life and unconventional marital
arrangement outshone the Nobel Prize and caused him considerable difficulty
in securing a tenured position until 1939, when he became the Director
of the Institute for Advanced Studies in Dublin. In the midst of the
turmoil of his tenure issues in 1935, during extensive correspondence
with Albert Einstein, he proposed what is now called the Schr\"{o}dinger's
cat thought experiment.

As one of the doctrines of the Copenhagen interpretation, a system
stops being in a superposition of states and becomes one or the other,
only when an observation takes place, which collapses the entire wave
function. Both Einstein and Schr\"{o}dinger thought this to be downright
preposterous.

In the same period of time during which Einstein presented his EPR
paper, he also wrote to Schr\"{o}dinger comparing the absurdity of the
Copenhagen interpretation to the notion of a keg of gun powder that
is simultaneously both exploded and unexploded. In an attempt to take
Einstein's thoughts a step further, the famed creator of wave mechanics
conjured up the enigmatic Schr\"{o}dinger's cat,
which
by virtue of being unobservable inside a box and subject to
a random quantum trigger that may or may not kill the kitty, is put into a quantum
superposition of being alive and dead at the same time.

Einstein was
duly impressed and wrote in a letter to Schr\"{o}dinger a bit later:
\begin{quote}
You are the only contemporary physicist, besides Laue, who sees that
one cannot get around the assumption of reality, if only one is honest.
Most of them simply do not see what sort of risky game they are playing
with reality---reality as something independent of what is experimentally
established.
They somehow believe that the quantum theory provides a description of reality, and even a \emph{complete} description; this 
interpretation is, however, refuted most elegantly
by your system of radioactive atom + Geiger counter + amplifier + charge of gun powder
+ cat in a box, in which the $\psi$-function of the system contains both
the cat alive and blown to bits. Nobody really doubts that the presence
or absence of the cat is something independent of the act of observation.\\
\cite[p.~39]{Einstein1967}
\end{quote}
Although Schr\"{o}dinger loved animals and apparently had a cat named
Milton at the time he devised his paradox, no one knows why he chose
a cat and not a dog or something else. Schr\"{o}dinger's thought experiment,
however, had little impact during his lifetime. From the time he proposed
it and till his death, it was scarcely mentioned in the literature.
Even Schr\"{o}dinger rarely brought it up. After its rather clandestine
appearance, Schr\"{o}dinger's cat went into a long slumber of nearly half
a century.

The absurd scenario portrayed in the thought experiment remained mostly
an academic curiosity until the 1980s, when it was realized that,
under suitable conditions, a macroscopic object with many microscopic
degrees of freedom could behave quantum mechanically, provided that
it was sufficiently decoupled from its environment. Such decoupling
is required because quantum superposition rapidly decoheres as a result
of complicated interactions with the environment and its inherently
irreversible thermal processes.

In the 1980s, there were also some other very exciting developments.
The French physicist Alain Aspect, in 1982, gave a
quite definitive experimental demonstration of quantum entanglement that led
to a burst of activities \cite{Aspect1982a,Aspect1982b}. In 1984, Charles Bennet and Gilles Brassard proposed a theoretical system for quantum cryptography using
photons in a superposition state to create a secure key distribution \cite{BB84}.
Achievement of a mesoscopic Schr\"{o}dinger's cat state allowing potential
studies of quantum decoherence generated considerable enthusiasm as well.

The explosion of publication of research articles on these subjects
is too voluminous to describe here. An extensive list of publications
on decoherence can be found in papers by its prominent investigator,
W.~H.~Zurek \cite{Zurek1,Zurek2,Zurek3,Zurek4}, as well as in the book by M.~Schlosshauer and his
other papers \cite{Schlosshauer1,Schlosshauer2,Schlosshauer3}. Only a selection of especially pertinent papers
on Schr\"{o}dinger's cat will now be discussed.

The physics Nobel Laureates Serge Haroche and David Wineland have
allowed Schr\"{o}dinger's cat to roam the most prestigious halls by mentioning
it in their Nobel lectures \cite{Nobel1,Nobel2}. Wineland shared the Nobel Prize with Haroche
in 2012 ``for ground breaking experimental methods
that enable measuring and manipulation of individual quantum systems.''
They independently developed methods for measuring and manipulating
individual particles while preserving their quantum mechanical nature
in ways that were previously thought unachievable. The two researchers
have taken different routes to the study of some of the same phenomena.

Wineland and his group used a trapped beryllium ion, laser
cooled to its zero-point energy \cite{Monroe1996}. A Schr\"{o}dinger's cat state was ingenuously
produced by applying a sequence of laser pulses to create superposition
of the ion's internal hyperfine electronic states with its external
positional states resulting from its zero-point motion in the potential
well of the trapping electro-magnetic field. A mesoscopic distance
of more than 80 nm separating the individual wave packets of about
7 nm was accomplished creating a mesoscopic Schr\"{o}dinger's cat state
that could allow controlled studies of quantum decoherence and study
of the quantum-classical boundary. Quantum decoherence has received
great interest to find a realistic solution of the long standing measurement
problem and more recently for application to quantum computing \cite{Zhang2017,Song2017} and quantum cryptography \cite{Srik2017}.

Of necessity, to measure the outcome of any quantum system we have
to use a macroscopic device composed of many quantum particles interacting
in an incredibly complicated way, and also partaking in irreversible
thermal processes. Is it then surprising that the transition from
the quantum to classical domain is the most difficult as well as a
controversial subject in quantum physics?

Fortunately, with the help of many sophisticated contemporary experiments
accompanied by highly developed theoretical investigations, significant
progress is being achieved although a very substantial difference
exists between theoretically anticipated and experimentally observed
values of decoherence time. A consensus seems to be developing in
favor of the environmental decoherence model pioneered by Zurek
\cite{Zurek1,Zurek2,Zurek3,Zurek4}. According to his theoretical analysis, the decoherence time
of a macroscopic object with a mass of 1~g at a temperature of
300~K extended over a distance of 1~cm could be as short as $10^{-23}$~s.
This, however, appears to be far too short compared to some
recent experimental investigations.

Observation has confirmed that quantum coherence is sustained
for longer than a few picoseconds inside photosynthetic light harvesting
pigment-protein complexes at room temperature \cite{photo1,photo2,photo3,photo4,photo5,photo6,photo7,photo8}. Even more spectacular
are analyses of recent experimental studies of the avian compass \cite{Gauger2011}.
Quantum superposition and entanglement are sustained in this living
system for at least tens of microseconds. Similar time frames for
sustaining quantum superposition have been conjectured in the tubulin
molecules of microtubules in the cytoskeletons of living
organisms \cite{Hameroff2014}.

These observations are strikingly at variance with the view that life
is too warm and wet for such quantum phenomena to endure over such
a relatively long time. It would appear that living entities may develop
at least in particular locations some sorts of shielding mechanism
to protect quantum superposition from environmental decoherence. But
this leaves us in a quandary about how long quantum superposition
could last in a living cat weighing at least 5~kg and containing some
$10^{27}$ atoms. In the absence of any meaningful data, the best
educated guess would be that decoherence will perhaps occur in extremely
short time scale under ambient conditions.

The coherence of superposition in large quantum objects is usually
demonstrated by a double slit type diffraction experiment. Such studies
have been performed with molecules as large as 1 nm, namely carbon
60 Bucky balls \cite{Arndt1999}. Recently, a diffraction pattern has been
observed in experiment of superposed molecules containing 10~000 atomic
mass units \cite{Eibenberger2013}.

Most remarkably, using a rather sophisticated procedure, coherent
superposition has been demonstrated in a macroscopic object containing
an estimated 10 trillion atoms \cite{OConnell2010}. For this purpose, the investigators
used a 40~micrometer long mechanical resonator, just large enough to be
visible with the naked eye. The resonator with a resonant frequency
of 6.175~GHz to its first excited phonon state was cooled to a temperature
of merely 25~mK over absolute zero and put in a very high vacuum to
minimize environmental decoherence. Under these circumstances, the
resonator was confirmed to be in its ground state. Then a signal from
a coupled qubit possessing the resonance frequency of 6.175~GHz was
injected into the resonator thereby transferring the superposition
feature of the qubit to the macroscopic object. Superposition of the
ground and the first excited phonon state of the macroscopic resonator
lasted for the resonator relaxation time of 6.1~ns.

The above demonstration provides strong evidence that quantum mechanics
and its attendant aspect of superposition applies to macroscopic objects
and can be revealed under appropriate circumstances. Can it apply
to Schr\"{o}dinger's cat? The answer in principle should be yes---at least
at an extremely short time scale. But to prove it, the cat will surely
perish for other reasons! Because in order to conduct the experiment,
it would be necessary to remove all sources of environmental decoherence
by exposing the cat to exceptionally low temperatures and high vacuum
and perhaps stopping irreversible metabolic processes as well.

\section{Conclusion}

By now the Schr\"{o}dinger's cat has been in existence for over eight
decades surviving longer than the mythical cat's nine lives. Considering
the fact that quantum superposition has been convincingly established
in quantum, mesoscopic and a macroscopic domain, Schr\"{o}dinger's cat
is likely to live on, symbolizing a profound truth that quantum reality
exists in all scales. We do not observe it in our daily macroscopic
world because it is masked for all practical purposes, predominantly
by environmental decoherence possessing irreversible thermal effects.

It should be noted however, that the concerns, which brought the Schr\"{o}dinger's
cat into existence are on their way out. We have cogently pointed
out that the probabilistic feature of a quantum particle is an inherent
aspect arising out of the genesis of the particle as a wave packet
comprising a collection of irregular disturbances of the underlying
quantum fields \cite{Bhaumik2015}. This is in addition to the innate uncertainty
relation $\Delta x\Delta p\geq\frac{1}{2}\hbar$  of a real wave packet.
Comprehension of the reality of the wave packet acting as a particle
was possible only after the development of the standard model in quantum field theory
and was thus unavailable to both Einstein and Schr\"{o}dinger
and to the framers of the Copenhagen interpretation.

Bohr's contention that there is no evocative reality before measurement
collapses the wave function was not entirely without merit. But his
emphatic denial of any reality whatsoever appears unjustifiable. His
emphasis on the critical role of a conscious observer for measurement
is also untenable, since the universe developed to a fairly mature
state obeying quantum rules long before the possibility of emergence
of conscious beings. Einstein's insistence on the existence of reality
in all scales from quantum to classical would be correct, if only
we allow him the understandable failure to see that the quantum reality
is not necessarily the same as the classical reality. Had the Bohr--Einstein
debate take place today, it probably would have been declared a draw!

\section*{Acknowledgement}

The author wishes to thank Joseph Rudnick and Zvi Bern for helpful
discussions.

\end{document}